\documentclass[a4paper,12pt]{article}%
\usepackage{amsmath}
\usepackage{amssymb}
\usepackage{amsfonts}
\usepackage[onehalfspacing,nodisplayskipstretch]{setspace}
\usepackage{graphicx}%
\usepackage{booktabs}
\usepackage{subcaption}
\usepackage{pgfplots}
\pgfplotsset{compat=1.18}
\providecommand{\U}[1]{\protect\rule{.1in}{.1in}}

\newtheorem{theorem}{Theorem}

\newtheorem{corollary}[theorem]{Corollary}

\newtheorem{proposition}[theorem]{Proposition}

\newtheorem{example}{Example}

\let\originalexample\example
\let\endoriginalexample\endexample

\renewenvironment{example}
  {\originalexample\normalfont}
  {\endoriginalexample}
  
\newenvironment{proof}[1][Proof]{\noindent\textbf{#1.} }{\ \rule{0.5em}{0.5em}}
\newcommand{\Rev}{\text{\normalfont\scshape Rev}}

\newcommand\ignore[1]{}
\newcommand{\ktm}[1]{#1_{\text{\normalfont\scshape menu}}}
\newcommand{\ktp}[1]{#1_{\text{\normalfont\scshape price}}}
\def\KTM{keep-the-menu}
\def\KTP{keep-the-pricing}

\begin{document}

\title{Extending Mechanisms\thanks{First version: June 2024.}}
\author{Sergiu Hart\thanks{The Hebrew University of Jerusalem (Federmann Center for
the Study of Rationality, Institute of Mathematics, and Department of
Economics).\quad\emph{E-mail}: \texttt{hart@huji.ac.il} \quad\emph{Web site}:
\texttt{https://ma.huji.ac.il/hart}}
\and Noam Nisan\thanks{The Hebrew University of Jerusalem (Federmann Center for the
Study of Rationality and School of Computer Science and Engineering).
\emph{E-mail}: \texttt{noam@cs.huji.ac.il} \quad\emph{Web site}:
\texttt{http://www.cs.huji.ac.il/\symbol{126}noam}}}
\maketitle

\begin{abstract}
Given an incentive-compatible mechanism for a single agent on a restricted domain of valuations, we  consider its incentive-compatible extensions to the entire valuation space. We identify 
two extreme extensions:
the \emph{\textquotedblleft keep-the-menu\textquotedblright}\ extension, which minimizes buyer payoffs and maximizes canonical prices pointwise, and the \emph{\textquotedblleft keep-the-pricing\textquotedblright}\ extension, which maximizes buyer payoffs and minimizes canonical prices pointwise.
\end{abstract}%

\def\@biblabel#1{#1\hfill}
\def\thebibliography#1{\section*{References}
\addcontentsline{toc}{section}{References}
\list{}{
\labelwidth0pt
\leftmargin1.8em
\itemindent-1.8em
\usecounter{enumi}}
\def\newblock{\hskip.11em plus .33em minus .07em}
\sloppy\clubpenalty4000\widowpenalty4000
\sfcode`\.=1000\relax\def\baselinestretch{1}\large\normalsize}
\let\endthebibliography=\endlist

\section{Introduction}

Suppose that you have a mechanism that is defined and incentive compatible on a subdomain of possible valuations (types) of the participants. Can it always
be extended to a mechanism that is incentive
compatible for all possible valuations of the participants? 

Let us look at an example in the simplest case of selling a single good to a single buyer, who is risk neutral and has quasi-linear preferences.  
Consider the mechanism in which the
buyer with valuation $0$ does not get the good (and pays nothing); the buyer with valuation
$20$ gets the good with probability $1/2$ and pays $5$; 
and the buyer with
valuation $30$ gets the good with certainty and pays  
$18$; see Figure \ref{fig:mechanism_table}.
This
mechanism is defined on the subdomain of valuations $\{0,20,30\}$.
On this
subdomain the mechanism is indeed incentive compatible as no buyer with 
one of these valuations can gain
by declaring any other valuation from this subdomain.  Now, can we define
an incentive-compatible mechanism for all
possible buyer valuations 
in $\mathbb{R}$
that agrees with this mechanism on the subdomain?  

\begin{figure}[htbp]
\centering

    \vspace{0pt}
    \setlength{\tabcolsep}{3pt}
    {\footnotesize
    \begin{tabular}{|c|c|c|c|}
        \hline
        \begin{tabular}{@{}c@{}}
            \textbf{buyer}\\
            \textbf{valuation}
        \end{tabular}
        &
        \begin{tabular}{@{}c@{}}
            \textbf{probability of}\\
            \textbf{getting the good}
        \end{tabular}
        &
         \begin{tabular}{@{}c@{}}
            \textbf{payment}\\
            \textbf{to seller}
        \end{tabular}
        &
         \begin{tabular}{@{}c@{}}
            \textbf{buyer}\\
            \textbf{payoff}
        \end{tabular}
     \\
        \hline\hline
        $0$  & $0$   & $0$  & $0$\\
        \hline
        $20$ & $1/2$ & $5$  &$5$\\
        \hline
        $30$ & $1$   & $18$ & $12$\\
        \hline
    \end{tabular}
    }
    \caption{Example of a mechanism defined on a subdomain.}
    \label{fig:mechanism_table}
\end{figure}
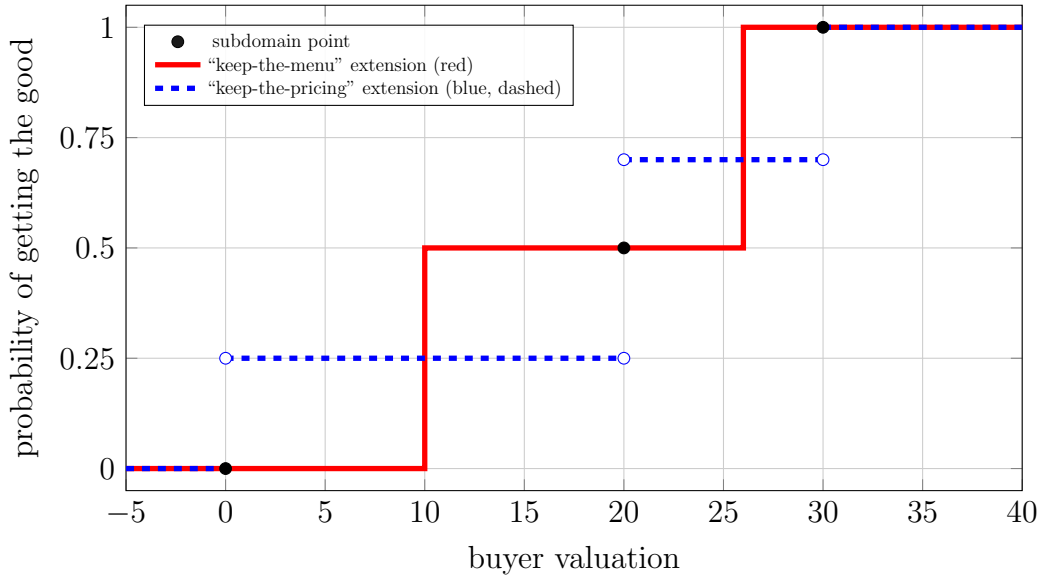
\begin{figure}[htbp]
    \centering
    \begin{tikzpicture}
        \begin{axis}[
            width=0.98\linewidth,
            height=8cm,
            xlabel={buyer valuation},
            ylabel={probability of getting the good},
            xmin=-5, xmax=40,
            ymin=-0.05, ymax=1.05,
            ytick={0,0.25,0.5,0.75,1},
            grid=both,
            grid style={
                line width=0.2pt,
                draw=gray!20
            },
            major grid style={
                line width=0.4pt,
                draw=gray!40
            },
            legend style={
                at={(0.02,0.96)},
                anchor=north west,
                nodes={scale=0.60, transform shape},
                fill=white,
                fill opacity=0.9,
                text opacity=1,
                cells={anchor=west}
            }
        ]

        \addplot[
            only marks,
            mark=*,
            mark size=2.2pt,
            fill=black,
            draw=black
        ] coordinates {
            (0,0)
            (20,0.5)
            (30,1)
        };
        \addlegendentry{\ subdomain point}

        \addplot[
            red,
            line width=2pt
        ] coordinates {
            (-5,0)
            (10,0)
            (10,0.5)
            (26,0.5)
            (26,1)
            (40,1)
        };
        \addlegendentry{``\KTM'' extension (red)}
        
        \addplot[
            blue,dashed,
            line width=2pt
        ] coordinates {
            (-5,0)
            (0,0)
        };
        \addlegendentry{``\KTP'' extension (blue, dashed)}
        
        \addplot[
            blue,dashed,
            line width=2pt,
            forget plot
        ] coordinates {
            (0,0.25)
            (20,0.25)
        };

        \addplot[
            blue,dashed,
            line width=2pt,
            forget plot
        ] coordinates {
            (20,0.7)
            (30,0.7)
        };

        \addplot[
            blue,dashed,
            line width=2pt,
            forget plot
        ] coordinates {
            (30,1)
            (40,1)
        };

        \addplot[
            only marks,
            mark=*,
            mark size=2.2pt,
            draw=blue,
            fill=white,
            forget plot
        ] coordinates {
            (0,0.25)
            (20,0.25)
            (20,0.7)
            (30,0.7)
        };

        \end{axis}
    \end{tikzpicture}
    \caption{Two extensions: allocation probability as a function of buyer valuation (black points specify the original allocations, open blue circles exclude endpoints, and vertical red segments indicate permissible tie-breaking allocations).}
    \label{fig:allocation_graph}

\end{figure}

At first sight,
requiring incentive compatibility for all possible buyer valuations introduces many new
constraints and it is not clear that this can be done consistently. However,
it is well known and usually taken for granted that by just offering the
menu of the possible outcomes, as defined for the given valuations in the subdomain,
and letting every buyer valuation choose among them,
one immediately gets an incentive-compatible mechanism for all possible
buyer valuations in\footnote{This observation 
holds for a single
participant, and does not imply that we can extend a {\em multi-participant} mechanism from an arbitrary
subdomain.  See, e.g., Lundy and Fu 2020.} $\mathbb{R}$. 
In our example if we offer the menu that was already given to us, ``the buyer can choose among getting the good with probability
$1/2$ at a price of $5$, getting it for sure at a price of $18$, and not getting it at all and not paying anything,'' one gets an incentive-compatible extension that is defined for all buyer valuations and its
allocation probabilities are shown in
Figure \ref{fig:allocation_graph} as the red line.

But are there other (incentive-compatible) extensions?
Figure \ref{fig:allocation_graph} also shows another possible extension
whose allocation probabilities are given by the blue dashed line.  (The payments are not shown on the graph but 
can be immediately derived.)  For example, in the ``blue extension,''
buyer valuations $20<x<30$ pay $9$ and get the good with probability $7/10$ (and for
$x=20$ the buyer pays $5$ and gets the good with probability $1/2$ and
for $x=30$ the buyer pays $18$ and gets the good for sure).  
The price $9$ for the allocation $7/10$ is the lowest price that preserves the choices of all original types. Indeed, at price $9$ types $20$ and $30$ are getting their original payoffs:
\[
\frac{7}{10} \cdot 20 -9 = 5, \quad\quad \frac{7}{10} \cdot 30 - 9 =12
\]
(type $0$ strictly prefers its original no-purchase option), and so any price less than $9$ would make types $20$ and $30$ switch to the new allocation $7/10$.
More generally, this extension offers every allocation at the lowest price that leaves the original
choices optimal.\footnote{The reader may wonder where the allocation $7/10$
comes from. Algebraically, requiring an allocation $g$ at
price $t$ to give types $20$ and $30$ their original payoffs
yields the equations $20g-t=5$ and $30g-t=12$, whose solution
is $g=7/10$ and $t=9$. Geometrically, $7/10$ is the slope of
the line joining the adjacent points $(20,5)$ and $(30,12)$ on the graph of the buyer payoff function of the original mechanism;
see Figure~\ref{fig:b} below.}
This example
is developed fully in Section \ref{sec:Examples} as part of our general framework which 
will, we hope, remove any mystery about how it is derived.

As mentioned, the ``red extension'' is obtained by offering the buyer 
the menu that was already defined in the subdomain; we call this the \emph{``\KTM''} extension. 
As we will see, the ``blue extension'' is obtained by offering 
each allocation at its ``canonical'' price in the original mechanism
(the canonical price of an allocation is the lowest price that leaves all prescribed choices optimal); we call this the \emph{``\KTP''} extension.
It yields the richest possible menu to the buyer. These two extensions turn out to be the pointwise \emph{minimal} and \emph{maximal}
extensions, respectively, in terms of the \emph{buyer's payoff} function.
Moreover, they correspond to, respectively, the pointwise \emph{maximal} 
and \emph{minimal canonical pricing} functions. 

These general results do not require the mechanism to be optimal for some distribution whose support is the given subdomain. Indeed, the mechanism above is not optimal for any random valuation with support \(\{0,20,30\}\) (the payment of type $20$ can be increased from $5$ to, say, $6$ while keeping incentive compatibility and individual rationality). Moreover, in the one-good case, we show that an optimal mechanism with finite positive revenue has a unique extension (see Appendix \ref{sus-a:opt1g}). 
With multiple goods, however, even a uniquely optimal restricted mechanism may have multiple extensions with different buyer payoff functions (see Example 2).

\section{Setup}

Our setup will be rather abstract (yet
standard; see Hart and Nisan 2012/2017;
2013/2019; 2025; 2026; Hart and Reny 2015). It has a single agent, and covers, in particular, a single seller selling multiple goods to
a single buyer with additive demand or with unit demand, and general
implementation problems with multiple options  (see Hart and Nisan 2026).
Our setting has a single risk-neutral quasi-linear buyer with $k$-dimensional valuation $x \in \mathbb{R}^{k}$.  
Let $\Gamma\subset\mathbb{R}^{k}$
be a fixed compact and convex \emph{set of allocations}, such as the unit cube
$[0,1]^{k}$ (which captures the setting of selling multiple goods to a buyer with an additive valuation), or a unit simplex $\{g \in \mathbb{R}^k_+:\sum_ig_i \le 1\}$ or $\{g \in \mathbb{R}^k_+:\sum_ig_i = 1\}$ (which capture the unit-demand and implementation settings), etc.
A mechanism on a domain $D\subseteq\mathbb{R}^k$ of valuations is a pair
$\mu=(q,s)$, where $q:D\rightarrow\Gamma$ is the \emph{allocation
function} and $s:D\rightarrow\mathbb{R}$ is the \emph{payment function}. 
A buyer
with valuation $x \in D$ receives the allocation 
$q(x) \in \Gamma$ and
pays $s(x)$ to the seller; his payoff is thus $b(x):=\sum_{i=1}^k q_i(x) x_i - s(x) = q(x) \cdot x - s(x)$, and we refer to $b:D \rightarrow \mathbb{R}$ as the \emph{buyer payoff function}.  

\emph{From now on, ``mechanism'' will always mean an incentive-compatible (IC) $\Gamma$-mechanism.}\footnote{\label{ftn:ir}To guarantee individual rationality (IR), 
include the zero valuation $x=(0,...,0)$ in $D_{0}$ and the zero allocation $g=(0,...,0)$ in $\Gamma$, and set $q_{0}(0,...,0)=(0,...,0)$ and $s_{0}(0,...,0)=0$; IR is then IC with respect to the zero valuation.}

Let
$\mu_{0}=(q_{0},s_{0})$ be a
mechanism
defined on a nonempty \emph{domain} $D_{0}%
\subseteq\mathbb{R}^{k}$ (for instance, $D_{0}$ is the support of a $k$-good
random valuation $X$); i.e., $(q_{0},s_{0}):D_{0}\rightarrow\Gamma
\times\mathbb{R}$. Let $Q_{0}:=q_{0}(D_{0})\subseteq\Gamma$ be the \emph{range
of allocations} of $\mu_{0}$. The \emph{buyer payoff function} $b_{0}%
:D_{0}\rightarrow\mathbb{R}$ is given by%
\[
b_{0}(x):=q_{0}(x)\cdot x-s_{0}(x)
\]
for every $x\in D_{0}$, and the \emph{pricing function} $p_{0}:Q_{0}%
\rightarrow\mathbb{R}$ by%
\[
p_{0}(g):=s_{0}(x)\text{\ \ for every }x\in D_{0}\text{ such that }q_{0}(x)=g
\]
for every $g\in Q_{0}$ (this is well defined by IC). The IC constraints
$b_{0}(x)=q_{0}(x)\cdot x-s_{0}(x)\geq q_{0}(y)\cdot x-s_{0}(y)$ for every
$x,y\in D_{0}$ are equivalently expressed as $b_{0}(x)\geq g\cdot x-p_{0}(g)$
for every $g\in Q_{0}$, or $p_{0}(g)\geq g\cdot x-b_{0}(x)$ for every $x\in
D_{0}$, with equality when $g=q_{0}(x)$. Equivalently, $b_{0}(x) + p_{0}(g)\geq g\cdot x$
for every $x \in D_0$ and $g\in Q_{0}$,  with equality when $g=q_{0}(x)$. Thus,%
\begin{align}
b_{0}(x)  & =\sup_{g\in Q_{0}}(g\cdot x-p_{0}(g))\text{\ \ for every }x\in
D_{0},\label{eq:b0}\\
p_{0}(g)  & =\sup_{x\in D_{0}}(g\cdot x-b_{0}(x))\text{\ \ for every }g\in
Q_{0}\text{.}\label{eq:p0}%
\end{align}

\section{Extensions}

We use the above formulas to extend $b_{0}$ and $p_{0}$ to their entire
spaces, $\mathbb{R}^{k}$ and $\Gamma$, respectively:\footnote{One may replace
$\mathbb{R}^{k}$ throughout with its nonnegative orthant $\mathbb{R}_{+}^{k}$;
see Remark (c) after the proof of Theorem \ref{th:main}.} we define
$\underline{b}:\mathbb{R}^{k}\rightarrow\mathbb{R}$ and $\underline{p}%
:\Gamma\rightarrow\bar{\mathbb{R}}:=\mathbb{R}\cup\{\infty\} $ by%
\begin{align}
\underline{b}(x)  &
{\;:=\;}%
\sup_{g\in Q_{0}}(g\cdot x-p_{0}(g))\;\;\text{for every }x\in\mathbb{R}%
^{k},\label{eq:bD}\\
\underline{p}(g)  &
{\;:=\;}%
\sup_{x\in D_{0}}(g\cdot x-b_{0}(x))\text{\ \ for every }g\in\Gamma
;\label{eq:pD}%
\end{align}
thus, $\underline{b}(x)=b_{0}(x)$ for every $x\in D_{0}$ and $\underline{p}%
(g)=p_{0}(g)$ for every $g\in Q_{0}$.

\bigskip

\noindent\textbf{Remark. }The function $\underline{b}$ is everywhere finite,
because it is finite on the nonempty set $D_{0}$ (where it coincides with
$b_{0}$), and, for $y\in\mathbb{R}^{k}$ and $x\in D_{0}$, we have
$\underline{b}(y)\leq\underline{b}(x)+\sup_{g\in Q_{0}}g\cdot(y-x)\leq
b_{0}(x)+\gamma\left\Vert y-x\right\Vert $, where $\gamma:=\max_{g\in\Gamma
}\left\Vert g\right\Vert $.

The function $\underline{p}$, however, may take infinite values, as shown in
Ben Moshe, Hart, and Nisan (2022). For example: there are two goods ($k=2$),
and $\mu_{0}$ sells the first good at a price of $1$; if $D_{0}$ contains
points $(0,x_{2})$ with arbitrarily large $x_{2}$ then the price of the second
good must be $\infty$.

The same applies to the functions $\overline{b}$ and $\overline{p}$ defined
below: $\overline{b}$ is everywhere finite, while $\overline{p}$ may take
infinite values.

\bigskip

A mechanism $\mu=(q,s):D\rightarrow\Gamma\times\mathbb{R}$ defined on a domain
$D\subseteq\mathbb{R}^{k}$ is a \emph{(mechanism-)extension} of $\mu_{0}$ if
$D\supseteq D_{0}$ and $\mu$ coincides with $\mu_{0}$ on $D_{0}$ (i.e.,
$(q,s)(x)=(q_{0},s_{0})(x)$ for every $x\in D_{0}$). Let $Q:=q(D)\supseteq
Q_{0}$ be the range of allocations of $\mu$, and let $b:D\rightarrow
\mathbb{R}$ and $p:Q\rightarrow\mathbb{R}$ be its buyer payoff and pricing
functions. Then for every $x\in D$ we have%
\begin{align*}
b(x)  & =\sup_{g\in Q}(g\cdot x-p(g))\geq\sup_{g\in Q_{0}}(g\cdot x-p(g))\\
& =\sup_{g\in Q_{0}}(g\cdot x-p_{0}(g))=\underline{b}(x)
\end{align*}
(the first equality is by (\ref{eq:b0}) applied to $\mu$). Similarly, for
every $g\in Q$ we have%
\begin{align*}
p(g)  & =\sup_{x\in D}(g\cdot x-b(x))\geq\sup_{x\in D_{0}}(g\cdot x-b(x))\\
& =\sup_{x\in D_{0}}(g\cdot x-b_{0}(x))=\underline{p}(g)
\end{align*}
(the first equality is by (\ref{eq:p0}) applied to $\mu$).

\bigskip

The functions $\underline{b}:\mathbb{R}^{k}\rightarrow\mathbb{R}$ and
$\underline{p}:\Gamma\rightarrow\bar{\mathbb{R}}$ thus provide \emph{lower
bounds} on $b$ and $p,$ respectively:%
\begin{align*}
b(x)  & \geq\underline{b}(x)\;\;\text{for every }x\in D,\\
p(g)  & \geq\underline{p}(g)\;\;\text{for every }g\in Q.
\end{align*}

\bigskip

We can then use the above inequalities to get \emph{upper bounds}, as follows.
Define $\overline{b}:\mathbb{R}^{k}\rightarrow\mathbb{R}$ and $\overline
{p}:\Gamma\rightarrow\bar{\mathbb{R}}$ by\footnote{The function $\overline{b}$ is finite since $\underline{p}(g) \ge g \cdot x - b_0(x)$ for a fixed $x \in D_0$ implies $\overline{b}(y) \le b_0(x) + \sup_{g \in \Gamma} g \cdot (y-x) \le b_0(x) + \gamma \left\Vert y-x\right\Vert$ for every $y \in \mathbb{R}^k$.}
\begin{align}
\overline{b}(x)  &
{\;:=\;}%
\sup_{g\in\Gamma}(g\cdot x-\underline{p}(g))\;\;\text{for every }%
x\in\mathbb{R}^{k},\label{eq:bU}\\
\overline{p}(g)  &
{\;:=\;}%
\sup_{x\in\mathbb{R}^{k}}(g\cdot x-\underline{b}(x))\text{\ \ for every }%
g\in\Gamma.\label{eq:pU}%
\end{align}
Then for every $x\in D$ we have%
\begin{align*}
b(x)  & =\sup_{g\in Q}(g\cdot x-p(g))\leq\sup_{g\in Q}(g\cdot x-\underline{p}%
(g))\\
& \leq\sup_{g\in\Gamma}(g\cdot x-\underline{p}(g))=\overline{b}(x),
\end{align*}
and for every $g\in Q$ we have%
\begin{align*}
p(g)  & =\sup_{x\in D}(g\cdot x-b(x))\leq\sup_{x\in D}(g\cdot x-\underline{b}%
(x))\\
& \leq\sup_{x\in\mathbb{R}^{k}}(g\cdot x-\underline{b}(x))=\overline{p}(g).
\end{align*}

Before stating the results, we recall the notion of the \textquotedblleft
canonical\textquotedblright\ extension of the pricing function (Hart and Reny
2015; see also Ben Moshe, Hart, and Nisan 2022), which assigns to each unused allocation
the lowest possible price that does not affect the choices in the mechanism.
Let $\mu=(q,s)$ be a $\Gamma$-mechanism with domain of valuations $D$ and
domain of allocations $Q=q(D)$, and let $p:Q\rightarrow\mathbb{R}$ be its
pricing function. Let $g\in\Gamma\backslash Q$ be an allocation that is not
used in $\mu$; what price $\tilde{p}(g)$ should be assigned to it so that no
valuation $x\in D$ would strictly prefer to pay $\tilde{p}(g)$ for the
allocation $g$ to its current choice? This requirement, which is $g\cdot
x-\tilde{p}(g)\leq b(x)$ for every $x\in D$, gives the condition $\tilde
{p}(g)\geq\sup_{x\in D}(g\cdot x-b(x)).$ The right-hand side, denoted
$p^{c}(g)$, is thus the minimal, or \emph{canonical}, price of $g$:%
\begin{equation}
p^{c}(g):=\sup_{x\in D}(g\cdot x-b(x))\label{eq:pc}%
\end{equation}
(for $g\in Q$ we have $p^{c}(g)=p(g)$ by (\ref{eq:p0}) applied to $\mu$). To
summarize: $\tilde{p}:\Gamma\rightarrow\bar{\mathbb{R}}$ is an extended
pricing function of $\mu$ if and only if $\tilde{p}(g)\geq p^{c}(g)$ for all
$g\in\Gamma$, with equality for $g\in Q$.

\section{Main Result}

We can now state the main result:

\begin{theorem}
\label{th:main}Let $\mu_{0}$ be 
a 
mechanism with domain of
valuations $D_{0},$ range of allocations $Q_{0},$ buyer payoff function
$b_{0},$ and pricing function $p_{0}.$ Let $\underline{b}\,,\overline
{b}:\mathbb{R}^{k}\rightarrow\mathbb{R}$ and $\underline{p}\,,\overline
{p}:\Gamma\rightarrow\bar{\mathbb{R}}$ be defined by (\ref{eq:bD}%
)-(\ref{eq:pU}), i.e.,%
\begin{align*}
\underline{b}(x)  &
{\;:=\;}%
\sup_{g\in Q_{0}}(g\cdot x-p_{0}(g))\text{\ \ for every }x\in\mathbb{R}^{k},\\
\underline{p}(g)  &
{\;:=\;}%
\sup_{x\in D_{0}}(g\cdot x-b_{0}(x))\text{\ \ for every }g\in\Gamma,\\
\overline{b}(x)  &
{\;:=\;}%
\sup_{g\in\Gamma}\,(g\cdot x-\underline{p}(g))\text{\ \ for every }%
x\in\mathbb{R}^{k},\\
\overline{p}(g)  &
{\;:=\;}%
\sup_{x\in\mathbb{R}^{k}}(g\cdot x-\underline{b}(x))\text{\ \ for every }%
g\in\Gamma.
\end{align*}
Then:

\begin{description}
\item[(i)] There is a mechanism $\ktm{\mu}$ on $\mathbb{R}^{k}$ that extends
$\mu_{0}$ with buyer payoff function $\underline{b}$ and canonical pricing
function $\overline{p}$, and a mechanism $\ktp{\mu}$ on $\mathbb{R}^{k}$ that
extends $\mu_{0}$ with buyer payoff function $\overline{b}$ and canonical
pricing function $\underline{p}$.

\item[(ii)] For every mechanism $\mu$ that extends $\mu_{0}$, with domain $D$,
range of allocations $Q$, buyer payoff function $b$, and pricing function $p$,
we have
\begin{align*}
& ~\underline{b}(x)\leq b(x)\leq\overline{b}(x)\text{\ \ for every }x\in D,\\
& ~\underline{p}(g)\leq p(g)\leq\overline{p}(g)\text{\ \ for every }g\in Q,
\end{align*}
with equalities for $x\in D_{0}$ and $g\in Q_{0}$.

\item[(iii)] The function $\underline{p}$ is the canonical pricing function of
$\mu_{0}$, i.e., 
\[
\underline{p}(g)=p_{0}^{c}(g)\text{\ \ for every }g\in\Gamma,
\]
and
the canonical pricing function $p^{c}$ of every mechanism $\mu$ that extends
$\mu_{0}$ satisfies\footnote{The inequality in (ii) holds only on $Q$ (where
$p$ is defined); here, it holds on the entire set $\Gamma$.}%
\[
\underline{p}(g)\leq p^{c}(g)\leq\overline{p}(g)\text{\ \ for every }%
g\in\Gamma.
\]
\end{description}
\end{theorem}

The mechanism\footnote{While the extremal buyer and canonical pricing functions are unique, the corresponding mechanisms are unique only up to tie-breaking when the buyer is indifferent (which does not affect the above functions).} $\ktm{\mu}$, with buyer payoff function $\underline{b}$ and
canonical pricing function $\overline{p}$, is the standard
\emph{\textquotedblleft \KTM\textquotedblright\ }extension of
$\mu_{0}$: every valuation chooses from the minimal possible set, namely, the
original menu $\{(q_{0}(x),s_{0}(x)):x\in D_{0}\}$ (or its closure when
necessary).\footnote{See the Appendix of Hart and Reny (2015).} Consequently,
$\underline{b}$ yields the lowest possible buyer payoff for every valuation
outside $D_{0}$. Equivalently, the price of each allocation $g$ outside $Q_{0}$ is
kept just high enough (specifically, given by $\overline{p}$), so that
\emph{no} valuation in $\mathbb{R}^{k}$ (not just those in $D_{0}$) would
strictly prefer this allocation $g$ to the original menu.

The mechanism $\ktp{\mu}$, with buyer payoff function $\overline{b}$ and
canonical pricing function $\underline{p}$, is the \emph{\textquotedblleft
\KTP\textquotedblright}\ extension of $\mu_{0}$: its canonical
pricing function is identical to the canonical pricing function of the
original mechanism $\mu_{0}.$ These prices are the absolutely lowest possible
for $\mu_{0}$ (anything strictly less would induce some valuation in $D_{0}$
to change its choice), and so the buyer payoff (given by
$\overline{b}$) of each valuation outside $D_{0}$ is the highest possible.

\bigskip

Writing $\ktm{b}$ and $\ktm{p}^{c}$ for the buyer payoff function and the
canonical pricing function of the mechanism $\ktm{\mu}$, and similarly for $\ktp{\mu}$, we thus have:%
\begin{align*}
\underline{b}(x)=\ktm{b}(x)~\leq & ~b(x)~  \leq~\ktp{b}(x)=\overline{b}%
(x)\text{\ \ for every }x\in D,\\
\underline{p}(g)=p_{0}^{c}(g)=\ktp{p}^{c}(g)~\leq & ~p^{c}(g)  \leq~\ktm{p}%
^{c}(g)=\overline{p}(g)~~\text{for every~}g\in\Gamma,
\end{align*}
for every extension $\mu$ of $\mu_{0}$ with domain $D$, buyer payoff function
$b$, and canonical pricing function $p^{c}$.

We emphasize that an inequality between pricing functions $p$ does not imply the same inequality between the resulting payments $s(x)$ of a fixed type $x$, because the allocations may change; for instance, in Example 1 below we have $\ktp{s}(x)=9 > 5 = \ktm{s}(x)$ for, say, $x=25$ (while $\ktp{p}^c \le \ktm{p}^c$).

\medskip

Finally, a function $b:D \rightarrow \mathbb{R}$ on a domain $D$ is $\Gamma$-\emph{convex} if at each point $x \in D$ it has a subgradient\footnote{A vector $v$ is a \emph{subgradient} of a function
$f:A\rightarrow\bar{\mathbb{R}}$ at a point $a_{0}\in A$ where $f$ is finite
if there is a constant $\beta$ such that $f(a)\geq v\cdot a+\beta$ for all
$a\in A$, with equality for $a=a_{0}$; equivalently, $f(a)\geq f(a_{0}%
)+v\cdot(a-a_{0})$ for all $a\in A$. Geometrically, this says that the
function $f$ is everywhere above the hyperplane with slope $v$ that equals
$f(a_{0})$ at $a_{0}.$} in $\Gamma$. We have

\begin{corollary}
Let $D_0 \subseteq D \subseteq \mathbb{R}^k$ and let $b:D \rightarrow \mathbb{R}$ be a $\Gamma$-convex function on $D$. Then $b$ is the buyer payoff function of an extension $\mu$ to $D$ of $\mu_0$ if and only if
\[
\underline{b}(x) \le b(x) \le \overline{b}(x) ~~~\text{for every}~ x \in D.
\]
Equivalently, this holds if and only if
\begin{align*}
&b(x) = b_0(x) ~~\text{for every}~ x \in D_0,~\text{and}
\\&b(x) \ge \underline{b}(x) ~~~\text{for every}~ x \in D.
\end{align*}
\end{corollary}

\begin{proof}
Necessity follows from Theorem~\ref{th:main}. For sufficiency, note first that $\underline{b} \le b \le \overline{b}$ on $D$ implies $b=b_0$ on $D_0$ (where $\underline{b}$ and $\overline{b}$ coincide with $b_0$). Second, from $b=b_0$ on $D_0$ and $b\ge\underline{b}$ on $D$ we get, for every $x\in D_0$ and $y\in D$,
\[
b(y)\ge\underline{b}(y)
\ge b_0(x)+q_0(x)\cdot(y-x)
=b(x)+q_0(x)\cdot(y-x).
\]
Hence the original allocation $q_0(x)$ is a supporting allocation for $b$ at each $x \in D_0$. Choosing supporting allocations in $\Gamma$ at the remaining points and setting $s(x)=q(x)\cdot x-b(x)$ gives the required extension.
\end{proof}

\medskip

As a consequence, whenever the two bounds $\underline{b}$ and $\overline{b}$ differ, there is a continuum of distinct extensions to the entire space (with buyer payoff functions $\lambda \underline{b} + (1-\lambda) \overline{b}$ for $\lambda \in [0,1]$; use the convexity of $\Gamma$).

\section{Examples}
\label{sec:Examples}

Before we get to the simple proof, we proceed with three examples.

\begin{example}
Selling one good: $k=1,$ $\Gamma=[0,1]$, and $\mu_{0}$ is
given by the table specified in the introduction:%
\[%
\begin{tabular}
[c]{|c|cc|c|}\hline
$x$ & $q_{0}(x)$ & $s_{0}(x)$ & $b_{0}(x)$\\\hline\hline
$0$ & $0$ & $0$ & $0$\\
$20$ & $\frac{1}{2}$ & $5$ & $5$\\
$30$ & $1$ & $18$ & $12$\\\hline
\end{tabular}
\;\;\;\;\;\;\;\;\;\;%
\begin{tabular}
[c]{|c|c|}\hline
$g$ & $p_{0}(g)$\\\hline\hline
$0$ & $0$\\
$\frac{1}{2}$ & $5$\\
$1$ & $18$\\\hline
\end{tabular}
\]
Thus, $D_{0}=\{0,20,30\}$ and $Q_{0}=\{0,\frac{1}{2},1\}.$ We then get (see
Figures \ref{fig:b} and \ref{fig:p}):

\[
\underline{b}(x)=\left\{
\begin{tabular}
[c]{lll}%
$0,$ &  & if $x\leq10,$\\
$\frac{1}{2}x-5,$ &  & if $10\leq x\leq26,$\\
$x-18,$ &  & if $26\leq x.$%
\end{tabular}
\right.
\]%
\[
\underline{p}(g)=\left\{
\begin{tabular}
[c]{lll}%
$0,$ &  & if $0\leq g\leq\frac{1}{4},$\\
$20g-5,$ &  & if $\frac{1}{4}\leq g\leq\frac{7}{10},$\\
$30g-12,$ &  & if $\frac{7}{10}\leq g\leq1.$%
\end{tabular}
\right.
\]%
\[
\overline{b}(x)=\left\{
\begin{tabular}
[c]{lll}%
$0$ &  & if $x\leq0,$\\
$\frac{1}{4}x,$ &  & if $0\leq x\leq20,$\\
$\frac{7}{10}x-9,$ &  & if $20\leq x\leq30,$\\
$x-18,$ &  & if $30\leq x.$%
\end{tabular}
\right.
\]%
\[
\overline{p}(g)=\left\{
\begin{tabular}
[c]{lll}%
$10g,$ &  & if $0\leq g\leq\frac{1}{2},$\\
$26g-8,$ &  & if $\frac{1}{2}\leq g\leq1.$%
\end{tabular}
\right.
\]

\begin{figure}[htbp]
    \centering
    \begin{subfigure}[t]{0.48\textwidth}
        \centering
        \includegraphics[width=\linewidth]{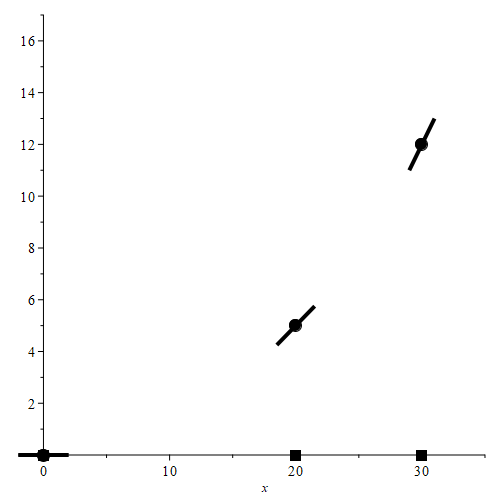}
        \caption{$b_0$ and the supporting lines, with slopes $q_0(x)$.}
    \end{subfigure}
    \hfill
    \begin{subfigure}[t]{0.48\textwidth}
        \centering
        \includegraphics[width=\linewidth]{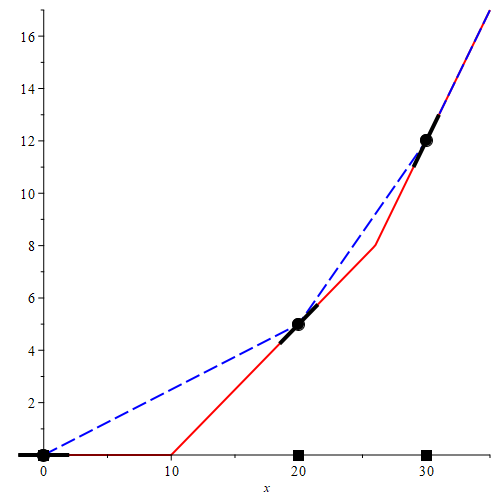}
        \caption{$\underline{b}$ in red, and $\overline{b}$ in blue (dashed).}
    \end{subfigure}

    \caption{%
        The buyer payoff function on the subdomain (left) and the two extreme extensions (right)
    }
    \label{fig:b}
\end{figure}

\begin{figure}[htbp]
    \centering
    \begin{subfigure}[t]{0.48\textwidth}
        \centering
        \includegraphics[width=\linewidth]{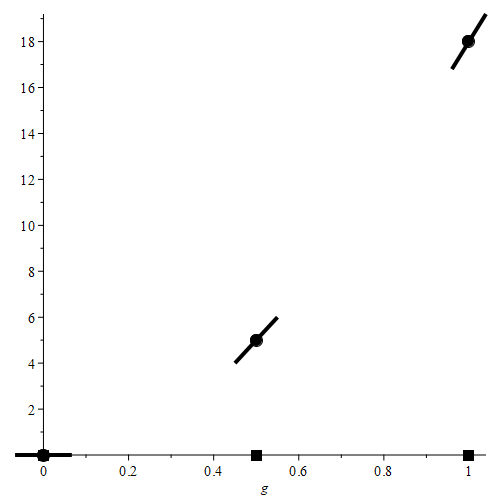}
        \caption{$p_0$ and the supporting lines, with slopes $x$ such that $q_0(x)=g$.}
    \end{subfigure}
    \hfill
    \begin{subfigure}[t]{0.48\textwidth}
        \centering
        \includegraphics[width=\linewidth]{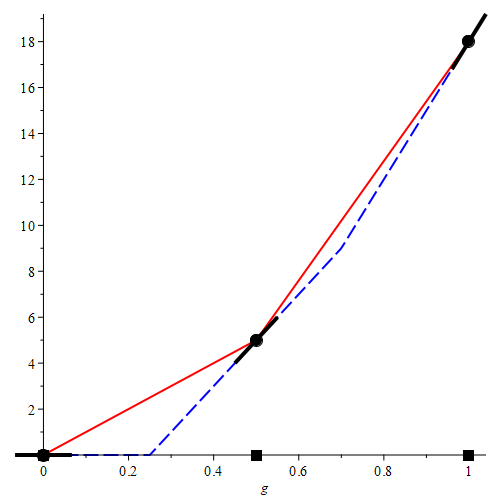}
        \caption{$\underline{p}$ in blue (dashed), and $\overline{p}$ in red.}
    \end{subfigure}

    \caption{%
        The pricing function on the subdomain (left) and the two extreme extensions (right)
    }
    \label{fig:p}
\end{figure}

The corresponding mechanisms, as depicted in Figure \ref{fig:allocation_graph}, are then obtained by setting 
$q(x)=b'(x)$ where the derivative $b'$ of $b$ is defined (which is
at almost every $x$) and an arbitrary
subgradient otherwise (with $q(x)=q_0(x)$ for types $x \in D_0$), and then  
$s(x)=q(x) \cdot x - b(x)$.  

The \KTM\ extension
whose allocation probability is depicted by the red line in Figure \ref{fig:allocation_graph} can be recovered from $\underline{b}$ 
(with its associated pricing $\overline{p}$):
$\ktm{q}(x) = 0$ for $x < 10$, 
$\ktm{q}(x) = 1/2$ for $10 < x < 26$,
and $\ktm{q}(x) = 1$ for $x > 26$, where at the transition points $x=10$ and $x=26$ 
one can arbitrarily choose any value\footnote{If one wants to literally keep the menu, then one should choose only one of the two extreme allocations.}
$0 \le \ktm{q}(10) \le 1/2$ and
$1/2 \le \ktm{q}(26) \le 1$.

The \KTP\ extension whose allocation probability is depicted
by the blue line in Figure
\ref{fig:allocation_graph} can be recovered from $\overline{b}$ 
(with its associated pricing $\underline{p}$):
$\ktp{q}(x)=0$ for $x < 0$,
$\ktp{q}(x)=1/4$ for $0 < x < 20$,
$\ktp{q}(x)=7/10$ for $20 < x < 30$,
and $\ktp{q}(x)=1$ for $x > 30$.  
Here, the transition points $x \in \{0, 20, 30\}$ are the subdomain points, and 
so for $\ktp{q}$ to extend $q_0$ we
must choose the subgradients to agree with
it: $\ktp{q}(0)=0$, $\ktp{q}(20)=1/2$, and $\ktp{q}(30)=1$. Thus, the \KTP\ mechanism is uniquely determined here.
\end{example}

\begin{example}\label{ex:2}
Selling $k \ge 2$ additive goods: $\Gamma=[0,1]^k$, the subdomain is $D_0 = \{(0,...,0), (1,...,1)\}$, and the mechanism $\mu_0$ sells the whole bundle $(1,...,1)$ to valuation $(1,...,1)$ at price $k$: 

\[%
\begin{tabular}
[c]{|c|cc|c|}\hline
$x$ & $q_{0}(x)$ & $s_{0}(x)$ & $b_{0}(x)$\\\hline\hline
$(0,...,0)$ & $(0,...,0)$ & $0$ & $0$\\
$(1,...,1)$ & $(1,...,1)$ & $k$ & $0$\\\hline
\end{tabular}
\;\;\;\;\;\;\;\;\;\;%
\begin{tabular}
[c]{|c|c|}\hline
$g$ & $p_{0}(g)$\\\hline\hline
$(0,...,0)$ & $0$\\
$(1,...,1)$ & $k$\\\hline
\end{tabular}
\]

We get
\begin{align*}
\underline{b}(x) & = \max\left\{\sum_i x_i -k,0\right\},
\\ 
\underline{p}(g) & =\sum_i g_i,
\\
\overline{b}(x) & =\sum_i \max\left\{x_i-1,0\right\},
\\
\overline{p}(g) & =
\left\{
\begin{tabular}
[c]{lll}%
$k\alpha$ &  & if $g =(\alpha,...,\alpha)$ for $\alpha \in [0,1]$,\\
$\infty$, &  & otherwise.\\
\end{tabular}
\right.
\end{align*}
Thus, the 
\KTM\ extension is the bundling mechanism with bundle price $k$, 
%
and the 
\KTP\ extension is the separate-selling mechanism with price $1$ for each good
(i.e., each good is sold at price $1$ independently of the other goods).

Now, $\mu_0$ extracts all the surplus (i.e., $s_0(x)=\sum_i x_i$ for every $x \in D_0$), and so it is the unique optimal mechanism for any random valuation $X$ with support $D_0$. This shows that for multiple goods, the mechanism being optimal, and even uniquely optimal, does not guarantee a unique extension (in contrast to the one-good case: 
see Appendix \ref{sus-a:opt1g}).

\end{example} 

\begin{example}\label{k}
Selling two additive goods: $k=2$, $\Gamma=[0,1]^{2}$, and
$\mu_{0}$ is the following variant of the nonmonotonic mechanism of Hart and
Reny (2015):%
\[%
\begin{tabular}
[c]{|c|cc|c|}\hline
$x$ & $q_{0}(x)$ & $s_{0}(x)$ & $b_{0}(x)$\\\hline\hline
$(0,0)$ & $(0,0)$ & $0$ & $0$\\
$(10,0)$ & $(1,0)$ & $10$ & $0$\\
$(0,20)$ & $(0,1)$ & $20$ & $0$\\
$(20,30)$ & $(1,1)$ & $40$ & $10$\\\hline
\end{tabular}
\;\;\;\;\;\;\;\;\;\;%
\begin{tabular}
[c]{|c|c|}\hline
$g$ & $p_{0}(g)$\\\hline\hline
$(0,0)$ & $0$\\
$(1,0)$ & $10$\\
$(0,1)$ & $20$\\
$(1,1)$ & $40$\\\hline
\end{tabular}
.
\]
Thus,
\begin{align*}
D_{0}&=\{(0,0),(10,0),(0,20),(20,30)\},\\
Q_{0}&=\{(0,0),(1,0),(0,1),(1,1)\},
\end{align*}
and we get%
\begin{align*}
\underline{b}(x) & =\max\{0,x_{1}-10,x_{2}-20,x_{1}+x_{2}-40\},
\\
\underline{p}(g)  & =\max\{0-0,10g_{1}-0,20g_{2}-0,20g_{1}+30g_{2}-10\}\\
& =\max\{10g_{1},20g_{2},20g_{1}+30g_{2}-10\}.
\end{align*}
To compute $\overline{b}$, the points that matter are $(g,\underline{p}(g))$
where $g$ is one of the four vertices of $[0,1]^{2}$ (i.e., $g\in Q_{0}$), and
in addition the point $g^{\#}=(2/5,1/5),$ where $\underline{p}(g^{\#})=4$
(which is an extreme point of the epigraph of $\underline{p}$); we get%
\[
\overline{b}(x)=\max\left\{  0,x_{1}-10,x_{2}-20,x_{1}+x_{2}-40,\frac{2}%
{5}x_{1}+\frac{1}{5}x_{2}-4\right\}  .
\]
Similarly, for $\overline{p}$ the points that matter are $(x,\underline{b}%
(x))$ where $x$ is one of the four points of $D_{0}$ (in fact, only
$x=(20,30)$ matters), and in addition the point $x^{\#}=(10,20),$ where
$\underline{b}(x^{\#})=0$ (which is an extreme point of the epigraph of
$\underline{b}$); we get%
\begin{align*}
\overline{p}(g)  & =\max\{10g_{1},20g_{2},20g_{1}+30g_{2}-10,10g_{1}%
+20g_{2}\}\\
& =\max\{20g_{1}+30g_{2}-10,10g_{1}+20g_{2}\}.
\end{align*}

Thus, $\underline{b}$ and $\overline{b}$ coincide everywhere except in the
region $\Delta$ where $g^{\#}$ is uniquely optimal, which is the interior of
the triangle with vertices $(10,0)$, $(0,20)$, and $(20,30)$. This implies that
every extension $b$ coincides with them outside $\Delta$ as well, and
so in particular \emph{every extension} to $\mathbb{R}^2$ (or $\mathbb{R}^2_+$) yields a \emph{nonmonotonic mechanism}: take for
instance the two points $y=(1,21)\ll(20,29)=z$ (which are outside the above
triangle), then we have $s(y)=20 > 10=s(z)$. In contrast, the payment function $s_0$ of the given mechanism $\mu_0$ is monotonic on its restricted domain $D_0$.
\end{example}

\section{Proof}\label{s:proof}

Before proving Theorem \ref{th:main}, we summarize the properties of the $b$
and $p$ functions in the following Proposition.

\begin{proposition}
\label{p:properties}Let $\mu_{0}$ be a
mechanism with domain of
valuations $D_{0},$ range of allocations $Q_{0},$ buyer payoff function
$b_{0},$ and pricing function $p_{0}.$ Let $\underline{b}\,,\overline
{b}:\mathbb{R}^{k}\rightarrow\mathbb{R}$ and $\underline{p}\,,\overline
{p}:\Gamma\rightarrow\bar{\mathbb{R}}$ be defined by (\ref{eq:bD}%
)-(\ref{eq:pU}). Then:

\begin{description}
\item[(i)] The functions $\underline{b}$ and $\overline{b}$ are convex
extensions of $b_{0}$ to $\mathbb{R}^{k}$, and satisfy $\underline{b}%
\leq\overline{b}$; similarly, the functions $\underline{p}$ and $\overline{p}$
are convex extensions of $p_{0}$ to $\Gamma$, and satisfy $\underline{p}%
\leq\overline{p}$:%
\[%
\begin{tabular}
[c]{cl}%
$\underline{b}(x)\leq\overline{b}(x)$ & $\text{for every }x\in\mathbb{R}^{k}%
,$\\
$\underline{b}(x)=b_{0}(x)=\overline{b}(x)$ & $\text{for every }x\in D_{0},$\\
$\underline{p}(g)\leq\overline{p}(g)$ & $\text{for every }g\in\Gamma,$\\
$\underline{p}(g)=p_{0}(g)=\overline{p}(g)$ & $\text{for every }g\in Q_{0}.$%
\end{tabular}
\]

\item[(ii)] The functions $\underline{b}$ and $\overline{p}$ are Fenchel
conjugate functions,\footnote{Define the pricing functions $\underline{p}$ and
$\overline{p}$ to be $\infty$ outside $\Gamma$.} and so are the functions
$\overline{b}$ and $\underline{p}$; i.e., in addition to (\ref{eq:bU}) and
(\ref{eq:pU}), we have%
\begin{align}
\underline{b}(x)  & =\sup_{g\in\Gamma}(g\cdot x-\overline{p}(g))\text{\ \ for
every }x\in\mathbb{R}^{k},\label{eq:bDD}\\
\underline{p}(g)  & =\sup_{x\in\mathbb{R}^{k}}(g\cdot x-\overline
{b}(x))\text{\ \ for every }g\in\Gamma.\label{eq:pDD}%
\end{align}

\item[(iii)] Let $x\in D_{0}$ and $g:=q_{0}(x)\in Q_{0}$; then $g$ is a
subgradient at $x$ of each one of the functions $b_{0}$,
$\underline{b}$, and $\overline{b}$, and $x$ is a subgradient at $g$ of each
one of the functions $p_{0}$, $\underline{p}$, and $\overline{p}$.
\end{description}
\end{proposition}

\begin{proof}
(i) The four functions are convex since they were defined as suprema of affine
functions. The function $\underline{b}$ coincides with $b_{0}$ on $D_{0}$ by
definition, and similarly $\underline{p}$ coincides with $p_{0}$ on $Q_{0}$.

The function $\overline{b}$ is everywhere above $\underline{b}$, because for
every $x\in\mathbb{R}^{k}$ we have
\[
\overline{b}(x)=\sup_{g\in\Gamma}(g\cdot x-\underline{p}(g))\geq\sup_{g\in
Q_{0}}(g\cdot x-\underline{p}(g))=\sup_{g\in Q_{0}}(g\cdot x-p_{0}%
(g))=\underline{b}(x)
\]
(we used $\underline{p}=p_{0}$ on $Q_{0}$). Similarly, $\overline{p}%
\geq\underline{p}$.

To see that the function $\overline{b}$ is an extension of $b_{0}$: let $x\in
D_{0}$; for every $g\in\Gamma$ we have $\underline{p}(g)\geq g\cdot
x-b_{0}(x)$ (by the definition of $\underline{p}$), and so $b_{0}(x)\geq
g\cdot x-\underline{p}(g)$; taking the supremum over $g\in\Gamma$ yields
$b_{0}(x)\geq\overline{b}(x)$, and so $b_{0}(x)\geq\overline{b}(x)\geq
\underline{b}(x)=b_{0}(x)$. Similarly, $\overline{p}=p_{0}$ on $Q_{0}$.

(ii) We show (\ref{eq:pDD}): for every $x\in\mathbb{R}^{k}$ and $g\in\Gamma$
we have $\overline{b}(x)\geq g\cdot x-\underline{p}(g)$ (from the definition
of $\overline{b}$), and so $\underline{p}(g)\geq g\cdot x-\overline{b}(x)$,
which yields $\underline{p}(g)\geq\sup_{x\in\mathbb{R}^{k}}(g\cdot
x-\overline{b}(x))$; for the converse inequality, $\underline{p}(g)=\sup_{x\in
D_{0}}(g\cdot x-b_{0}(x))=\sup_{x\in D_{0}}(g\cdot x-\overline{b}(x))\leq
\sup_{x\in\mathbb{R}^{k}}(g\cdot x-\overline{b}(x))$. Similarly for
(\ref{eq:bDD}).

(iii) We have $b_{0}(x)=g\cdot x-p_{0}(g).$ For every $y\in\mathbb{R}^{k}$ it
then follows that $\overline{b}(y)\geq\underline{b}(y)\geq g\cdot y-p_{0}(g)$
(by (i) and the definition of $\underline{b}$), with equality at $y=x$ (by
(i)), and so $g$ is a subgradient at $x$ of both $\overline{b}$ and
$\underline{b}$, and thus also of their restriction $b_{0}$ to $D_{0}$.
Similarly, $x$ is a subgradient at $g$ of $\overline{p}$, $\underline{p}$, and
$p_{0}$.
\end{proof}

\bigskip

We can now prove Theorem \ref{th:main}.

\begin{proof}
[Proof of Theorem \ref{th:main}](i) Let $\mathcal{B}_{\Gamma}$ be the set of
all $\Gamma$\emph{-convex} functions on $\mathbb{R}^k$: convex functions $b:\mathbb{R}%
^{k}\rightarrow\mathbb{R}$ that have a subgradient in
$\Gamma$ at every $x$, i.e., $\partial b(x)\cap\Gamma\neq\emptyset$ (see Hart and Nisan
2026; note that here we do not require that $b(0,...,0)=0$). It is well
established (Rochet 1985; see also Hart and Nisan 2026) that $b\in
\mathcal{B}_{\Gamma}$ if and only if there is a $\Gamma$-mechanism $\mu$ with
buyer payoff function $b$: the allocation $q(x)$ is a subgradient of $b$ at
$x$. Now the functions $\underline{b}$ and $\overline{b}$ are each defined as
a supremum of a collection of affine functions with slopes in $\Gamma$ and are
everywhere finite, and so they are in\footnote{Use Theorem 23.4 and Corollary
13.1.1 of Rockafellar (1970) to show that all the subgradients are in $\Gamma
$. When the extension is to $\mathbb{R}_{+}^{k}$ rather than to the entire
space $\mathbb{R}^{k}$ (see Remark (c) below) there are additional
subgradients at boundary points of $\mathbb{R}_{+}^{k}$ (specifically: if
$x_{i}=0$ then one can arbitrarily decrease the $i$th coordinate of any
subgradient at $x$).} $\mathcal{B}_{\Gamma}$. 
To make $\mu_{i}=(q_{i},s_{i})$,
for $i \in \{\textsc{menu},\textsc{price}\}$, an extension of $\mu_{0}$, set $(q_{i},s_{i})(x):=(q_{0}%
,s_{0})(x)$ for every $x\in D_{0}$ (this is possible since, by Proposition
\ref{p:properties} (iii), $q_{0}(x)$ is a subgradient at $x$ of both
$\underline{b}$ and $\overline{b}$). The canonical pricing functions are
$\ktm{p}^{c}=\overline{p}$ by (\ref{eq:pU}) and $\ktp{p}^{c}=\underline{p}$ by
(\ref{eq:pDD}).

(ii) These inequalities were established in the text above.

(iii) The definitions of $\underline{p}$ and $p_{0}^{c}$ (see (\ref{eq:pD}),
and (\ref{eq:pc}) for $\mu_{0}$) are identical. Next, for the canonical
pricing function $p^{c}$ of an extension $\mu$ of $\mu_{0}$, for every
$g\in\Gamma$ we have, first,%
\[
p^{c}(g)=\sup_{x\in D}(g\cdot x-b(x))\leq\sup_{x\in D}(g\cdot x-\underline{b}%
(x))\leq\sup_{x\in\mathbb{R}^{k}}(g\cdot x-\underline{b}(x))=\overline{p}(g)
\]
(we used $b\geq\underline{b}$ by (ii) and $D\subseteq\mathbb{R}^{k}$); and
second,%
\[
\underline{p}(g)=\sup_{x\in D_{0}}(g\cdot x-b_{0}(x))=\sup_{x\in D_{0}}(g\cdot
x-b(x))\leq\sup_{x\in D}(g\cdot x-b(x))=p^{c}(g)
\]
(we used $b=b_{0}$ on $D_{0}$ and $D_{0}\subseteq D$).
\end{proof}

\bigskip

\noindent\textbf{Remarks. }\emph{(a)} By Proposition \ref{p:properties} (i)
and (iii), the functions $\underline{b}$ and $\overline{b}$ are
\textquotedblleft value-and-subgradient\textquotedblright\ extensions of
$b_{0}$, in the sense that they preserve not only the values $b_{0}(x)$ but
also the subgradients $q_{0}(x)$, for all $x\in D_{0}$. Moreover, among all
$\Gamma$-convex (i.e., in $\mathcal{B}_{\Gamma}$) value-and-subgradient
extensions of $b_{0}$, the function $\underline{b}$ is the (pointwise) minimal
one, and the function $\overline{b}$ is the (pointwise) maximal one. See
Figure \ref{fig:b} for the one-good Example 1 above. Similarly, $\underline{p}%
$ and $\overline{p}$ are the (pointwise) minimal and (pointwise) maximal,
respectively, closed\footnote{A convex function $\tilde{p}:\Gamma
\rightarrow\bar{\mathbb{R}}$ is \emph{closed} (lower semicontinuous) if
$\tilde{p}(g)\leq\liminf_{h\rightarrow g,h\in\Gamma}\tilde{p}(h)$ at every
$g\in\Gamma$. The functions
$\underline{p}$ and $\overline{p}$ are closed as suprema of continuous
(affine) functions.} convex value-and-subgradient extensions of $p_{0}$: they
preserve the values $p_{0}(g)$ and the subgradients $x$ for which $q_{0}%
(x)=g$, for all $g\in Q_{0}$. Proposition \ref{p:extreme} in Appendix \ref{sus-a:b-p}
provides a precise statement of these minimality and maximality results.
Geometrically, $\underline{b}$ is the upper envelope of the
prescribed supporting affine functions (with slopes given by the subgradients), whereas $\overline{b}$ is the largest $\Gamma$-convex extension of the prescribed values (the ``$\Gamma$-convexification'' of $b_0$).

\emph{(b) }A strengthening of Proposition \ref{p:properties} (iii): For $x\in
D_{0}$ and $g\in Q_{0}$, the following are all equivalent: $g$ is optimal for
$x\in D_{0}$ in $\mu_{0}$ (i.e., $b_{0}(x)=g\cdot x-p_{0}(g)$, which holds in
particular for $g=q_{0}(x)$); $g$ is a subgradient at $x$ of any of the
functions $b_{0}$, $\underline{b}$, or $\overline{b}$; and $x$ is a
subgradient at $g$ of any of the functions $p_{0}$, $\underline{p}$, or
$\overline{p}$ (the proof of Proposition \ref{p:properties} (iii) above holds
for any optimal $g$; the rest is straightforward).

\emph{(c) }One may check that replacing throughout $\mathbb{R}^{k}$ with its
nonnegative orthant $\mathbb{R}_{+}^{k}$ does not affect the results (some of
the functions may change). The Fenchel conjugacy in Proposition
\ref{p:properties} (ii) is now with respect to $\mathbb{R}_{+}^{k}$ and
$\Gamma$ only, and the maximality and minimality of $\overline{p}$ and
$\underline{p}$ are within the class of closed nondecreasing convex
value-and-subgradient extensions of $p_{0}$ (the functions $\underline{p}$ and
$\overline{p}$ are nondecreasing since they are suprema of affine functions of
the form $g\cdot x-b(x)$ with $x\in\mathbb{R}_{+}^{k}$, which are thus
nondecreasing; see the Remark in Appendix \ref{sus-a:b-p} for the modification needed in
the proof of Proposition \ref{p:extreme} (ii)).

\bigskip

Finally, when the optimal mechanism for a random valuation $X$ with support
$D_{0}$ (e.g., a finite support) is unique, this means that its restriction
$\mu_{0}$ to $D_{0}$ is unique ($X$-almost surely). Among mechanisms defined
on the entire space, there can be many whose restriction to $D_{0}$ is
$\mu_{0}$, and hence all of them are optimal for\footnote{Assume for simplicity that incentive compatibility entails individual rationality, cf. footnote \ref{ftn:ir}.} $X$.

\appendix{}

\section{Appendix}

\subsection{The Extreme Buyer Payoff and Pricing Functions}\label{sus-a:b-p}

We show here the minimality and maximality properties of the corresponding $b
$ and $p$ functions (see Remark (a) after the proof of Theorem \ref{th:main}).

\begin{proposition}
\label{p:extreme}(i) Let $\tilde{b}:\mathbb{R}^{k}\rightarrow\mathbb{R}$ be a
$\Gamma$-convex function that is a value-and-subgradient extension of $b_{0}$
(i.e., $\tilde{b}(x)=b_{0}(x)$ and $q_{0}(x)\in\partial\tilde{b}(x)$ for every
$x\in D_{0}$). Then $\underline{b}\leq\tilde{b}\leq\overline{b}.$

(ii) Let $\tilde{p}:\Gamma\rightarrow\bar{\mathbb{R}}$ be a closed convex
function that is a value-and-subgradient extension of $p_{0}$ (i.e.,
$\tilde{p}(g)=p_{0}(g)$ and $x\in\partial\tilde{p}(g)$ for every $x\in D_{0}$
with $q_{0}(x)=g$). Then $\underline{p}\leq\tilde{p}\leq\overline{p}$.
\end{proposition}

\begin{proof}
(i) Upper bound: Let $y\in\mathbb{R}^{k}$; since $\tilde{b}$ is $\Gamma
$-convex there is $g\in\partial\tilde{b}(y)\cap\Gamma$, and thus for every
$x\in D_{0}$ we have $b_{0}(x)=\tilde{b}(x)\geq\tilde{b}(y)+g\cdot(x-y)$, and
so $\tilde{b}(y)\leq g\cdot y-(g\cdot x-b_{0}(x)).$ This holds for every $x\in
D_{0},$ and thus $\tilde{b}(y)\leq g\cdot y-\sup_{x\in D_{0}}(g\cdot
x-b_{0}(x))=g\cdot y-\underline{p}(g)\leq\overline{b}(y)$.

Lower bound: For every $g\in Q_{0}$ let $x\in D_{0}$ be such that $g=q_{0}(x)
$, then $g\in\partial\tilde{b}(x)$, and so $\tilde{b}(y)\geq\tilde
{b}(x)+g\cdot(y-x)=g\cdot y-(g\cdot x-b_{0}(x))=g\cdot y-p_{0}(g).$ This holds
for every $g\in Q_{0}$, and thus $\tilde{b}(y)\geq\sup_{g\in Q_{0}}(g\cdot
y-p_{0}(g))=\underline{b}(y).$

(ii) Upper bound: Extend $\tilde{p}$ to $\mathbb{R}^{k}$ by putting $\tilde
{p}(g)=\infty$ for $g\notin\Gamma.$ Denoting the Fenchel conjugate of $f$ by
$f^{\ast}$, for every $x\in\mathbb{R}^{k}$ we have $\tilde{p}^{\ast}%
(x)=\sup_{g\in\mathbb{R}^{k}}(g\cdot x-\tilde{p}(g))\geq\sup_{g\in Q_{0}%
}(g\cdot x-\tilde{p}(g))=\sup_{g\in Q_{0}}(g\cdot x-p_{0}(g))=\underline{b}%
(x).$ Thus $\tilde{p}^{\ast}\geq\underline{b},$ and so $\tilde{p}=\tilde
{p}^{\ast\ast}\leq\underline{b}^{\ast}=\overline{p}$ (the first equality by
Theorem 12.2 of Rockafellar 1970, since $\tilde{p}$ is closed; the inequality
since conjugacy reverses order; and the second equality by (\ref{eq:pU})).

Lower bound: Let $h\in\Gamma$; for every $x\in D_{0}$ let $g:=q_{0}(x),$ then
$x\in\partial\tilde{p}(g),$ and so $\tilde{p}(h)\geq\tilde{p}(g)+(h-g)\cdot
x=h\cdot x-(g\cdot x-p_{0}(g))=h\cdot x-b_{0}(x).$ This holds for every $x\in
D_{0},$ and thus $\tilde{p}(h)\geq\sup_{x\in D_{0}}(h\cdot x-b_{0}%
(x))=\underline{p}(h).$
\end{proof}

\bigskip

\noindent\textbf{Remark.} The proof of the inequality $\tilde{b}\leq
\overline{b}$ does not use the subgradient-extension property (only the
value-extension property), and that of the inequality $\tilde{b}%
\geq\underline{b}$ does not use the $\Gamma$-convexity property. Similarly,
$\tilde{p}\leq\overline{p}$ does not use the subgradient-extension property,
and $\tilde{p}\geq\underline{p}$ does not use the closedness of $\tilde{p}$.
When dealing with $\mathbb{R}_{+}^{k}$ instead of $\mathbb{R}^{k}$ (see Remark
(c) following the proof of Theorem \ref{th:main}) one needs to adapt the proof
of $\tilde{p}\leq\overline{p}$: before using conjugacy, extend $\tilde{p}$ to
the nondecreasing closed convex function $\tilde{P}$ on $\mathbb{R}^{k}$ given
by $\tilde{P}(h):=\inf_{h\leq g\in\Gamma}\tilde{p}(g) $; then $\tilde{P}%
^{\ast}(x)=\infty$ for $x\notin\mathbb{R}_{+}^{k}$ and $\tilde{P}^{\ast
}(x)\geq\sup_{g\in Q_{0}}(g\cdot x-p_{0}(g))=\underline{b}(x)$ for
$x\in\mathbb{R}_{+}^{k},$ and so $\tilde{p}(g)=\tilde{P}(g)=\sup
_{x\in\mathbb{R}_{+}^{k}}(g\cdot x-\tilde{P}^{\ast}(x))\leq\sup_{x\in
\mathbb{R}_{+}^{k}}(g\cdot x-\underline{b}(x))=\overline{p}(g)$ for every $g\in{\Gamma}$.

\subsection{One-Good Optimal Mechanisms Have Unique Extensions}\label{sus-a:opt1g}

Consider the case of one good, i.e., $k=1$. A mechanism $\mu _{0}$ on $D_{0}$
is \emph{optimal} for a random valuation $X$ with support\footnote{%
The \emph{support} of $X$ is the smallest closed set $D_{0}$ such that $%
\mathbb{P}\left[ X\in D_{0}\right] =1$.} $D_{0}$ if it maximizes the revenue
among all IC and IR mechanisms on $D_{0}$. We will show here that optimal one-good mechanisms have a \emph{unique} extension. 

However, this result holds \emph{only for one good}: for multiple goods there are in general multiple extensions\footnote{We emphasize that the multiplicity here involves
distinct buyer payoff functions $b$, not merely different
tie-breaking choices for a fixed $b$. In the one-good case,
we prove below the uniqueness of
the extension \emph{mechanism}, i.e., its allocations and
payments at every valuation.} of an optimal mechanism, even if the mechanism is the unique optimal mechanism. Indeed, Example \ref{ex:2} above is such an instance: 
$\mu_0$, which 
is the unique optimal mechanism for any random valuation $X$ with support $D_0$, has $\underline{b} \neq \overline{b}$, and thus multiple extensions. Thus, for more than one good, even unique optimality does not yield a unique extension as in the one-good case.

\begin{figure}[!htpb]
\centering
\includegraphics[width=0.6\linewidth]{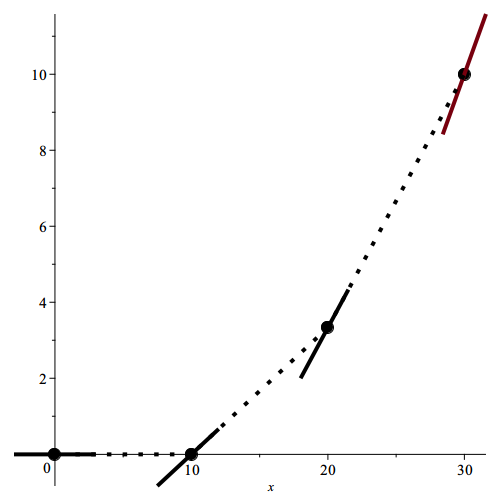}
        \caption{$b_0$ and $q_0$ for an optimal one-good mechanism.}
        \label{fig:unique}
\end{figure}

To see why the extension is unique for one dimension, let $\mu_0$ be an optimal mechanism for a random valuation $X$ with support  $D_0 \subset \mathbb{R}$, which we assume for now to be finite and consist of the points $x_1 < x_2 < ... <x_n$ in $\mathbb{R}$. It is easy to see\footnote{Nothing in this section relies on the Myerson (1981) characterization of optimal mechanisms as (convex combinations of) optimal pricing mechanisms.} that if $\mu_0$ is optimal then each valuation $x_i$ (for $i \ge 2$) is indifferent between its choice $(q_0,s_0)(x_i)$ and the choice $(q_0,s_0)(x_{i-1})$ of the valuation $x_{i-1}$ just below it (see (\ref{eq:ic(y,z)}) in the proof below). Geometrically, this means that the point $(x_i,b_0(x_i))$ on the graph of $b_0$ lies on the line tangent to $b_0$ at $x_{i-1}$ with slope $q_0(x_{i-1})$. By the value-and-subgradient condition, the buyer payoff function $b$ of any extension $\mu$ of $\mu_0$ must then coincide with this tangent in the interval $[x_{i-1},x_i]$ (because it is a convex function that coincides with $b_0$ at the two end points and it is weakly above the tangent). See Figure \ref{fig:unique} (and compare it with Figure~\ref{fig:b}, where there are multiple extensions, because $(x_i,b_0(x_i))$ is \emph{strictly} above the tangent at $x_{i-1}$).
In multiple dimensions, binding IC constraints may link types in different directions, and there is no analogous linear ordering of types.

For simplicity assume that $0\in D_{0}$ and $\mu _{0}$ satisfies $q_{0}(0)=0$
and $s_{0}(0)=0$, and so IR is just IC with respect to $0$.

\begin{theorem}
Let $k=1$ and $\Gamma =[0,1]$. Let $X$ be a random valuation with support $%
D_{0}$ that contains $0$, and whose optimal revenue $\Rev (X)$ satisfies $0< \Rev (X) < \infty$.
Let $\mu _{0}$ be a mechanism on $D_{0}$ with $q_{0}(0)=s_{0}(0)=0$ that is
optimal for $X$. Then $\mu _{0}$ has a unique extension to the entire space $%
\mathbb{R}$.
\end{theorem}

\begin{proof}
Let $\mu =(q,s)$ be an extension of $\mu _{0}$ to the entire space $\mathbb{R%
}$ (its existence is guaranteed by our main result), and so its buyer payoff
function $b$ is a value-and-subgradient $\Gamma $-convex (and thus
nondecreasing and nonexpansive) extension of $b_{0}$. Since $D_{0}$ is
closed, its complement $D_{0}^{C}:=\mathbb{R}\backslash D_{0}$ is an open
set, and thus the union of open bounded intervals $(y,z)$ with $y,z\in D_{0}$%
, together with, possibly, two open half-lines $(-\infty ,y)$ and $(z,\infty
)$ with $y,z\in D_{0}.$ We will show that $b(x)$ is uniquely determined on $%
D_{0}^{C}$ and affine on each such connected component of $D_{0}^{C}$;
consequently $q(x)=b^{\prime }(x)$ and $s(x)=q(x)x-b(x)$ are uniquely
determined there as well.

Consider first an interval $(y,z)\subseteq D_{0}^{C}$ with $y<z$ and $y,z\in
D_{0}$. By IC we have $b_{0}(z)=q_{0}(z)z-s_{0}(z)\geq q_{0}(y)z-s_{0}(y)$.
We claim that the optimality of $\mu _{0}$ implies equality, i.e.,%
\begin{equation}
b_{0}(z)=q_{0}(z)z-s_{0}(z)=q_{0}(y)z-s_{0}(y),  \label{eq:ic(y,z)}
\end{equation}
since otherwise increasing the payments of all $x\geq z$ in $D_{0}$ by the
same sufficiently small $\varepsilon >0$ preserves all IC and IR constraints
and strictly increases the revenue.\footnote{%
This is a standard argument: if the slack in the above inequality is larger
than $\varepsilon $, then every IC constraint of $z^{\prime }\geq z$ against 
$y^{\prime }\leq y$ has slack larger than $\varepsilon $ (by IC and the
monotonicity of $q_{0}$).} Geometrically, (\ref{eq:ic(y,z)}) says that the
point $(z,b_{0}(z))$ lies on the line through $(y,b_{0}(y))$ with
slope $q_{0}(y)$, and so $b$ must coincide with this
line on the interval $[y,z]$; i.e., $b(x)=L(x)$ for every $x\in [y,z]$, where $L$ is the affine
function $L(x):=q_{0}(y)x-s_{0}(y)$. Indeed, since $b(y)=b_{0}(y)=L(y)$
and $b(z)=b_{0}(z)=L(z)$, and $L$ is affine and $b$ is convex, we have, first, $%
b(x)\leq L(x)$ for every $x$ in $[y,z]$; and second, $b(x)\geq L(x)$ for every $x
$ by the value-and-subgradient condition at $y\in D_{0}$. Thus $b(x)=L(x),$
and so $q(x)=b^{\prime }(x)=L^{\prime }(x)=q_{0}(y)$, for every $x$ in $(y,z)
$.

Next, assume that $(z,\infty )\subseteq D_{0}^{C}$, and so $z\in D_{0}$ is
the maximal point in $D_{0}$. Then $q_{0}(z)=1$ (otherwise the revenue,
which we assumed finite and positive, can be increased: multiply $q_{0}$ and 
$s_{0}$ by $1+\varepsilon $ for small enough $\varepsilon >0$ so that $%
(1+\varepsilon )q_{0}(z)\leq 1$, which preserves IC and IR and multiplies
the revenue by $1+\varepsilon $), and thus for every $x$ we have $b(x)\geq
q_{0}(z)x-s_{0}(z)=x-s_{0}(z)$ (by the value-and-subgradient condition at $%
z\in D_{0}$), and for $x>z$ we have $b(x)\leq
b(z)+(x-z)=z-s_{0}(z)+x-z=x-s_{0}(z)$ (by nonexpansiveness). Thus $%
b(x)=x-s_{0}(z)$ and its derivative $b^{\prime }(x)=q(x)=1$ are uniquely
determined for all $x>z.$

Finally, assume that $(-\infty ,y)\subseteq D_{0}^{C}$, and so $y\in D_{0}$
is the minimal point in $D_{0}$. Then $b_{0}(y)=0$ (because $b_{0}(y)\geq 0$
by IR, and $b_{0}(y)\leq b_{0}(0)=0$ since $y\leq 0\in D_{0}$ and $b_{0}$ is
nondecreasing). Therefore $b(x)\leq b(y)=b_{0}(y)=0$ for every $x<y$
(because $b$ is nondecreasing), and $b(x)\geq q_{0}(0)x-s_{0}(0)=0$ (by the
value-and-subgradient condition at $0\in D_{0}$). Thus $b(x)=0$ and $%
q(x)=b^{\prime }(x)=0$ are uniquely determined for all $x<y.$

Thus $b$ and $q=b^{\prime }$ (and thus $s(x)=q(x)x-b(x)$) are uniquely
determined on the complement $D_{0}^{C}$ of $D_{0};$ since on $D_{0}$ they
are given by $\mu _{0}$, the uniqueness of $\mu $ has been established.
\end{proof}

\medskip

The proof shows the uniqueness of the extension of the optimal $\mu_0$ to \emph{any} domain $D$ that contains $D_0$.

\end{document}